\documentclass[12pt,a4paper]{article}

\usepackage{amssymb}
\usepackage{amscd}

\newcommand{\dd}{\mathrm{d}}
\newcommand{\vect}{\mathbf}
\newcommand{\Lied}[1]{\mathcal{L}_{\vect{#1}}}
\newcommand{\OmegaEL}{\Omega_{\mathrm{EL}}}
\newcommand{\at}[1]{\big|_{#1}}

\newcommand{\XS}{\mathcal{X}_{\mathrm{S}}}
\newcommand{\XH}{\mathcal{X}_{\mathrm{H}}}
\newcommand{\tr}{\mathrm{tr\,}}

\newtheorem{thm}{Theorem}
\newtheorem{cor}{Corollary}[thm]
\newtheorem{lem}{Lemma}

\newcommand{\comments}[1]{{}}
\newcommand{\tedious}[1]{}
\newcommand{\definition}{\emph}
\newcommand{\omits}[1]{}

\title{\bf The Euler-Lagrange Cohomology\\
  and General Volume-Preserving Systems}
\author{Bin Zhou\thanks{Email: zhoub@ihep.ac.cn}\\
  \small{Institute of High Energy Physics,
  Chinese Academy of Sciences,} \\
  \small{P. O. Box 918-4, Beijing 100039, P. R. China,}\\
  Han-Ying Guo\thanks{Email: hyguo@itp.ac.cn}\\
  \small{Institute of Theoretical Physics,
  Chinese Academy of Sciences,} \\
  \small{P. O. Box 2735, Beijing 100080, P. R. China,}\\
  Jianzhong Pan\thanks{Email: pjz@amss.ac.cn}\\
  \small{Academy of Mathematics and System Sciences, Chinese Academy of Sciences,} \\
  \small{Beijing 100080, P. R. China,}\\
  and\\
  Ke Wu\thanks{Email: wuke@itp.ac.cn}\\
  \small{Department of Mathematics, Capital Normal University,} \\
  \small{Beijing 100037, P. R. China}}

\date{\small\it (Draft ~~April 7, 2003)}

\begin{document}

\maketitle

\begin{abstract}
We briefly introduce the conception on Euler-Lagrange cohomology
groups on a symplectic manifold $(\mathcal{M}^{2n}, \omega)$ and
systematically present the general form of volume-preserving
equations on the manifold from the cohomological point of view. It
is shown that for every volume-preserving flow generated by these
equations there is an important 2-form that plays the analog role
with the Hamiltonian in the Hamilton mechanics. In addition, the
ordinary canonical equations with Hamiltonian $H$ are included as
a special case with the 2-form $\frac{1}{n-1}\,H\,\omega$. It is
studied the other volume preserving systems on $({\cal M}^{2n},
\omega)$. It is also explored the relations between our approach
and Feng-Shang's volume-preserving systems as well as the Nambu
mechanics.


\end{abstract}
\newpage
\tableofcontents

\newpage
\section{Introduction}
\label{sect:Intro}

As is well known, \omits{ the classical mechanics is well
established (see, for example, \cite{Arnold}, \cite{AM}). In the
classical mechanics, the systems with potential can be described
by both the Hamiltonian and Lagrangian formalism.}in the
Hamiltonian mechanics (see, for example, \cite{Arnold},
\cite{AM}), the canonical equations with Hamiltonian $H$ read
\begin{equation}
  \dot{q}^{\,i} = \frac{\partial H}{\partial p_i}, \qquad
  \dot{p}_i = - \frac{\partial H}{\partial q^i}.
\label{Heqs}
\end{equation}
\omits{For an autonomous system, the Hamiltonian $H$ manifestly
independent of the time $t$ is a function on the phase
space.}\omits{ $\mathcal{M}$, the cotangent bundle of the
configuration space.} In fact, a Hamiltonian system  defined on a
symplectic manifold $(\mathcal{M}^{2n},\omega)$ as its phase space
\omits{where $\mathcal{M}$ is a $2n$-dimensional orientable
differentiable manifold endowed} with a symplectic 2-form
$\omega$\omits{For the details of Hamiltonian mechanics on a
symplectic manifold, see, for example, \cite{Arnold}. According to
Darboux's theorem,} \omits{for each point $x$ in $\mathcal{M}$,
there is a coordinate neighborhood $(U; q^1,
\dots,q^n,p_1,\dots,p_n)$ of $x$ such that\footnote{ Repeated
indices are always assumed to be summed over 1 to $n$ unless
otherwise stated. For convenience, symbols such like
$\omega\at{U}$ are always abbreviated as $\omega$ later. }and
locally we always have
\begin{equation}
  \omega = \dd p_i\wedge\dd q^i, \qquad i=1, \cdots, n.
\label{omegadef}
\end{equation}
Thus a Hamiltonian system 
on $(\mathcal{M}^{2n},\omega)$ }can always be described locally by
eqs.~(\ref{Heqs}). The solutions of eqs.~(\ref{Heqs}) are the
integral curves of the Hamiltonian vector field
\begin{equation}
  \vect{X}_H := \frac{\partial H}{\partial p_i}
    \frac{\partial}{\partial q^i}
  - \frac{\partial H}{\partial q^i}\frac{\partial}{\partial p_i}
\end{equation}
on $\mathcal{M}^{2n}$. It is well known that the above Hamiltonian
system preserves both the symplectic form $\omega$ and the volume
form
\begin{eqnarray}\label{vform}
\tau:=\frac{1}{n!}\, \omega^n,
\end{eqnarray}
where $\omega^n$ is
the $n$-fold wedge product of $\omega$.

In the symplectic geometry \cite{ST}, for a vector field $\vect X$
on $\mathcal{M}^{2n}$, there is always a 1-form $E_{\vect X}:= -
i_{\vect{X}}\omega $, where $i_{\vect{X}}$ denotes the
contraction. The vector is symplectic, if $E$ is closed.
 On the other
hand, for a Hamiltonian vector field $\vect{X}_H$ on
$\mathcal{M}^{2n}$, the 1-form $E_{\vect{X}_H}$ is exact and the
ordinary canonical equations (\ref{Heqs}) can be expressed as
\begin{equation}
  - i_{\vect{X}_H}\omega = \dd H.
\end{equation}
Thus, a cohomology
 may be introduced :
\begin{equation}
  H_{\mathrm{EL}}:=\{E_{\vect{X}}|\dd E_{\vect{X}}=0\}/
  \{E_{\vect{X}}|E_{\vect{X}}=\dd \alpha\}.
\label{eq:HEL}
\end{equation}
In \cite{ELcoh2, ELcoh4}, in order to study its time discrete
version in discrete mechanics including the symplectic algorithm,
this cohomology has been introduced and is called the
Euler-Lagrange cohomology. It is simple but significant to see
that by definition the canonical equations (\ref{Heqs}) are in the
image of the cohomology. In addition, it is also straightforward
to show that the Euler-Lagrange cohomology is isomorphic to the de
Rahm cohomology on $\mathcal{M}^{2n}$ \cite{gpwz1, gpwz2} and that
both $\omega$ and $\tau$ in (\ref{vform}) are preserved not only
along the phase flow defined by the eqs (\ref{Heqs}) but also
along the symplectic flow. Namely,
\begin{eqnarray}
  \Lied{X_{\mathnormal{H}}}\omega = 0,&&\qquad
  \Lied{X_{\mathnormal{H}}}\tau=0,\\[1mm]
  \Lied{X_{\mathnormal{S}}}\omega = 0,&&\qquad
  \Lied{X_{\mathnormal{S}}}\tau=0.
\end{eqnarray}
 \omits{and
\begin{equation}
  \Lied{X_{\mathnormal{H}}}\tau \propto
  \Lied{X_{\mathnormal{H}}}(\omega^n)
  = n\,(\Lied{X_{\mathnormal{H}}}\omega)\wedge\omega^{n-1}=0,
\end{equation}
respectively.}

However, it should be emphasized  that there exist certain
important mechanical systems that preserve the volume of the phase
space only and cannot be transformed into the ordinary Hamiltonian
systems. For example, let us consider the following kind of linear
systems
\begin{equation}
  \ddot{q}^{\ i} = - k_{ij} \,q^j
\label{ls}
\end{equation}
on $\mathbb{R}^{n}$ with constant coefficients $k_{ij}$.
Obviously, eqs. (\ref{ls}) can be turned into the form:
\begin{equation}
  \dot{q}^{\ i} = \frac{\partial H}{\partial p_i}, \qquad
  \dot{p}_i = - \frac{\partial H}{\partial q^i} - a_{ij}\, q^j
\label{lsys}
\end{equation}
with
\begin{eqnarray}
  & & H = \frac{1}{2}\,\delta^{ij}\,p_i p_j
  + \frac{1}{2}\, s_{ij}\,q^i q^j,
\label{lsysH} \\
  & & s_{ij} = \frac{1}{2}\,( k_{ij} + k_{ji}), \qquad
  a_{ij} = \frac{1}{2}\,(k_{ij} - k_{ji}).
\label{lsyscoef}
\end{eqnarray}
When one of $a_{ij}$ is nonzero, 
the system (\ref{lsys}) is not an ordinary Hamiltonian system on
$\mathbb{R}^{2n}$.  In fact, the corresponding vector field
\begin{displaymath}
  \vect{X}_a = \vect{X}_H - a_{ij}\,q^j\,
  \frac{\partial}{\partial p_i}
\end{displaymath}
of (\ref{lsys}) is not even a symplectic vector field since the
Lie derivative
\begin{displaymath}
  \Lied{X_a}\omega = a_{ij}\,\dd q^i\wedge\dd q^j\neq 0 .
\end{displaymath}
However, it is easy to verify that the  system (\ref{lsys}) always
preserves the volume form of the phase space, i.e.
$\Lied{X_a}(\tau) = 0$.

It is worth while to mention that 
even some conservative system can be transformed as such kind of
non-Hamiltonian linear systems. Consider a system consisting of
two 1-dimensional linearly coupled oscillators with different
mass:
\begin{displaymath}
  m_1\,\ddot{q}^{\,1} = -k\,(q^2 - q^1), \qquad
  m_2\,\ddot{q}^{\,2} = -k\,(q^1 - q^2).
\end{displaymath}
Obviously, such a system does satisfy Newton's laws. Let $k_{11} =
-k/m_1$, $k_{12} = k/m_1$, $k_{21} = k/m_2$ and $k_{22} = -k/m_2$.
Then it is a system as described by eqs.~(\ref{ls}), with $i,j
=1,2$. When $m_1 \neq m_2$, we have such a system
that $k_{12} \neq k_{21}$.

The  volume-preserving systems are very important. This can also
be seen as follows. If a system $S$ on $(\mathcal{M}^{2n},\omega)$
is not volume-preserving, it can be extended as a
volume-preserving system $S'$ on
$(\mathcal{M}^{2n}\times\mathbb{R}^2, \omega')$ in such a way that
the orbits of $S$ are precisely the projection of the orbits of
$S'$ onto $\mathcal{M}^{2n}$. As a demonstration, let $q^i$ and
$p_i$ ($i = 1, \cdots, n$) be the Darboux coordinates on
$\mathcal{M}^{2n}$ and $q^0$, $p_0$  the ones on $\mathbb{R}^2$.
Then take $\omega' = \dd p_\mu\wedge \dd q^\mu = \omega + \dd
p_0\wedge\dd q^0,~~\mu=0,1,\cdots, n,$ as the symplectic structure
on $\mathcal{M}^{2n}\times\mathbb{R}^2$. Note that $\omega$ in the
expression should be written as $\pi^*\omega$ in which $\pi:
\mathcal{M}^{2n}\times\mathbb{R}^2 \longrightarrow
\mathcal{M}^{2n}$ is the projection. But, as a demonstration, we
do not need to give a rigorous description, although it can be
made.

Suppose that for the system $S$ on $\mathcal{M}^{2n}$,
$\Lied{X}(\omega^n) = D\, \omega^n$ where $D = D(q^i, p_i)$ is a
function and
$$\vect{X} = Q^i(q^1,\ldots,q^n,p_1,\ldots,p_n)\,
  \frac{\partial}{\partial q^i}
 + P_i(q^1,\ldots,q^n,p_1,\ldots,p_n)\,
 \frac{\partial}{\partial p_i}$$
is the vector field of $S$. Then a system $S'$ on
$\mathcal{M}\times\mathbb{R}^2$ corresponding to
$$\vect{X}' = Q^i\,\frac{\partial}{\partial q^i}
  + P_i\,\frac{\partial}{\partial p_i}
  - D(q^1,\ldots,q^n,p_1,\ldots,p_n)q^0 \,
  \frac{\partial}{\partial q^0}$$
can be constructed. It can be easily checked that $S'$ is a
volume-preserving system. And the orbits of $S$ are just the
projections of the orbits of $S'$. If all the properties of system
$S'$ are known, so the properties of $S$.

In this paper, we generalize the Euler-Lagrange cohomology to
higher forms on a symplectic manifold $(\mathcal{M, \omega})$ and
present a general form of equations that generate a
volume-preserving flow  from the cohomological point of view. It
is shown that for every volume-preserving flow there is a 2-form
that plays an analog role with the Hamiltonian  in the Hamilton
mechanics. The ordinary Hamiltonian equations with Hamiltonian $H$
are included as a special case with a 2-form
$\frac{1}{n-1}\,H\,\omega$.

The main results on the Euler-Lagrange cohomology will be quoted
and outlined without detailed proof. Some of the general content
and results can be found in \cite{gpwz1}
 and more details as well as extended topics will be given in
the forthcoming paper \cite{gpwz2}. In what follows, we first
generalize the Euler-Lagrange cohomology to higher order ones in
section 2. Then we present the general volume preserving equations
from cohomological point of view in section 3. We also show that
the canonical equations in the Hamilton mechanics and the kind of
linear systems just mentioned are special cases of the general
volume preserving equations. In section 4, we compare our general
volume-preserving equations with other volume-preserving systems
presented before. Especially, we explore the relations with
Feng-Shang's volume-preserving systems in their volume-preserving
algorithm\cite{FS} and with Nambu's mechanics\cite{Nambu}.
Finally, we end with some conclusion and remarks.


\section{The Euler-Lagrange Cohomology on Symplectic Manifolds}

In order to  present the general equations for volume-preserving
systems from the cohomological point of view, we need first to
generalize the Euler-Lagrange cohomology from $H_{\mathrm{EL}}$ of
1-forms to higher forms. In this way a sequence of Euler-Lagrange
cohomology groups $H^{2k-1}_{\mathrm{EL}}(\mathcal{M},\omega)$ can
be introduced. Then we consider the trivial class of the highest
one with $k = n$. The general volume-preserving equations will be
picked up from it. This is, in fact, a direct generalization for
the ordinary Hamilton system, where the canonical equations can be
represented by the 1-forms in the trivial class of
$H_{\mathrm{EL}}$ as defined in eq.~(\ref{eq:HEL}).

Briefly speaking, a volume-preserving system characterized by a
volume-preserving vector field, corresponds to a unique closed
$(2n-1)$-form on $\mathcal{M}^{2n}$ and \textit{vise versa}. Such
a system that corresponds to an exact $(2n-1)$-form can be
determined then by a 2-form up to at least a closed
2-form.


In this section, the concept of Euler-Lagrange cohomology groups
$H^{2k-1}_{\mathrm{EL}}(\mathcal{M},\omega)$ for $k = 1, \ldots,
n$ will be briefly introduced. Some of their important properties
will be quoted. But the details on the Euler-Lagrange cohomology
groups will be explained in the forthcoming paper \cite{gpwz2}.

For $k = 1,\ldots,n$,\omits{ where $n =
\frac{1}{2}\,\dim\mathcal{M} \geqslant 1$,} we may define the sets
\begin{eqnarray*}
  \XS^{2k-1}(\mathcal{M}^{2n},\omega) &
  := & \{\,\vect{X}\in{\mathcal X}({\cal M}^{2n})\,|\,
  \Lied{X}(\omega^k) = 0 \,\}, \\
  \XH^{2k-1}(\mathcal{M}^{2n},\omega) &
  := & \{\,\vect{X}\in{\cal X}({\cal M}^{2n})\,|\,
  - i_{\vect{X}}(\omega^k) \textrm{ is exact}\}.
\end{eqnarray*}
For simplicity, hereafter $\mathcal{X(M)}$ denotes the space of
smooth vector fields on $\mathcal{M}$ of dimension $2n$.  The
vector fields in $\XS^1(\mathcal{M},\omega)$ and
$\XH^1(\mathcal{X},\omega)$ are known as symplectic vector fields
and Hamiltonian vector fields, respectively. Those in
$\XS^{2n-1}(\mathcal{M},\omega)$ are just volume-preserving vector
fields. It is easy to prove that
\begin{eqnarray*}
  & & \XS^1(\mathcal{M},\omega) \subseteq \ldots \subseteq
  \XS^{2k-1}(\mathcal{M},\omega) \subseteq \XS^{2k+1}(\mathcal{M},\omega)
  \subseteq \ldots \subseteq\XS^{2n-1}(\mathcal{M},\omega), \\
  & & \XH^1(\mathcal{M},\omega) \subseteq \ldots \subseteq
  \XH^{2k-1}(\mathcal{M},\omega) \subseteq \XH^{2k+1}(\mathcal{M},\omega)
  \subseteq \ldots \subseteq\XH^{2n-1}(\mathcal{M},\omega)
\end{eqnarray*}
and
\begin{displaymath}
  \XH^{2k-1}(\mathcal{M},\omega) \subseteq \XS^{2k-1}(\mathcal{M},\omega).
\end{displaymath}
In fact, $\XS^{2k-1}(\mathcal{M},\omega)$ is a Lie algebra\omits{
under the commutation bracket of vector fields,} and
$\XH^{2k-1}(\mathcal{M},\omega)$ is an ideal of
$\XS^{2k-1}(\mathcal{M},\omega)$ because
\begin{displaymath}
  [\XS^{2k-1}(\mathcal{M},\omega),\XS^{2k-1}(\mathcal{M},\omega)]
  \subseteq \XH^{2k-1}(\mathcal{M},\omega).
\end{displaymath}

The quotient Lie algebra
\begin{equation}
  H^{2k-1}_{\mathrm{EL}}(\mathcal{M},\omega)
  :=
  \XS^{2k-1}(\mathcal{M},\omega)/\XH^{2k-1}(\mathcal{M},\omega)
\label{eq:HELdef}
\end{equation}
is  called the $2k-1$-st \definition{Euler-Lagrange cohomology
group}\cite{gpwz1},\cite{gpwz2},\cite{zgw}. It is called a group
because its Lie algebra structure is trivial: For each $k = 1$,
$\ldots$, $n$, $H^{2k-1}_{\mathrm{EL}}(\mathcal{M},\omega)$ is an
Abelian Lie algebra. For $n=1$,
$H^1_{\mathrm{EL}}(\mathcal{M},\omega)$ is the Euler-Lagrange
cohomology. And it can be proved that
$H^1_{\mathrm{EL}}(\mathcal{M},\omega)$ is linearly isomorphic to
the first de~Rham cohomology group
$H^1_{\mathrm{dR}}(\mathcal{M})$ as was mentioned in the
introduction.

For $k = n$, $\XS^{2n-1}(\mathcal{M},\omega)$ is the Lie algebra
of volume-preserving vector fields, including all the other Lie
algebras mentioned  above as its Lie subalgebras. Similarly to the
case of $k = 1$, the $2n-1$-st Euler-Lagrange cohomology group
$H^{2n-1}_{\mathrm{EL}}(\mathcal{M},\omega)$ is isomorphic to the
$(2n - 1)$-st de~Rham cohomology group
$H^{2n-1}_{\mathrm{dR}}(\mathcal{M})$. To prove this, let us first
introduce a lemma:

\begin{lem}
  For each integer $k=1,\ldots, n$ $(n \geqslant 1)$ and $x\in\mathcal{M}$, the
linear map $\nu_k: T_x \mathcal{M} \longrightarrow \Lambda_{2k-1}(T^*_x
\mathcal{M})$ defined by $\nu_k(\vect{X}) = - i_{\vect{X}}(\omega^k)$ is
injective. That is, $i_{\vect{X}}(\omega^k) = 0$ iff $\vect{X} = 0$.
\label{lem:lemma1}
\end{lem}

The Euler-Lagrange cohomology groups can also be
defined as
\begin{displaymath}
  H^{2k-1}_{\mathrm{EL}}(\mathcal{M},\omega)
  = \{ \theta\in\OmegaEL^{2k-1}(\mathcal{M},\omega)\,|\,\dd\theta=0\}/
  \{ \theta\in\OmegaEL^{2k-1}(\mathcal{M},\omega)\,|\,\theta
  \textrm{ is exact} \},
\end{displaymath}
where $\OmegaEL^{2k-1}(\mathcal{M},\omega) := \{ -i_{\vect{X}}(\omega^k) \,|\,
\vect{X}\in\mathcal{X}(\mathcal{M})\}.$ Using the above lemma, it is easy to
see that this definition is equivalent to (\ref{eq:HELdef}).

Since $\dim T_x\mathcal{M} = \dim \Lambda_{2n-1}(T^*_x\mathcal{M})$, lemma
\ref{lem:lemma1} implies that $\nu_n: T_x\mathcal{M}\longrightarrow
\Lambda_{2n-1}(T^*_x\mathcal{M})$ is a linear isomorphism, which induces a
linear isomorphism
$\nu_n:\mathcal{X(M)}\longrightarrow\Omega^{2n-1}(\mathcal{M})$
defined by
\begin{equation}
  \nu_n(\vect{X}) = - i_{\vect{X}}(\omega^n)
  = -n\,(i_{\vect{X}}\omega)\wedge\omega^{n-1}.
\end{equation}
Hence we obtain
\begin{thm}
The linear map $\nu_n:
\mathcal{X(M)}\longrightarrow\Omega^{2n-1}(\mathcal{M})$ is an
isomorphism. Under this isomorphism,
$\XS^{2n-1}(\mathcal{M},\omega)$ and
$\XH^{2n-1}(\mathcal{M},\omega)$ are isomorphic to
$Z^{2n-1}(\mathcal{M})$ and $B^{2n-1}(\mathcal{M})$, respectively.
\label{thm:thm1}
\end{thm}
In the above theorem, $Z^{2n-1}(\mathcal{M})$ is the space of closed
$(2n-1)$-forms on $\mathcal{M}$ and $B^{2n-1}(\mathcal{M})$ is the space of
exact $(2n-1)$-forms on $\mathcal{M}$. As a consequence, we obtain the
following corollary:
\begin{cor}
  The $2n-1$-st Euler-Lagrange cohomology group
$H^{2n-1}_{\mathrm{EL}}(\mathcal{M},\omega)$ is linearly
isomorphic to $H^{2n-1}_{\mathrm{dR}}(\mathcal{M})$, the
$(2n-1)$-st de~Rham cohomology group.
\end{cor}

When $\mathcal{M}$ is closed\omits{(i.e., compact and without boundary)},
$H^{2n-1}_{\mathrm{EL}}(\mathcal{M},\omega)$ is linearly
isomorphic to the dual space of
$H^1_{\mathrm{EL}}(\mathcal{M},\omega)$, because
  $H^{2k-1}_{\mathrm{dR}}(\mathcal{M})
  \cong (H^{2(n-k)+1}_{\mathrm{dR}}(\mathcal{M}))^*$
for such a manifold (see, for example, \cite{Warner}).
If $\mathcal{M}$ is not compact, this relation cannot be assured.

When $n > 2$, there is another lemma:
\begin{lem}
Let $n>2$. Then, for an arbitrary integer $k = 1, \ldots, n-2$ and
a point $x\in\mathcal{M}$, $\alpha\in\Lambda_2(T^*_x\mathcal{M})$
satisfies $\alpha\wedge\omega^k = 0$ iff $\alpha = 0$.
\label{lem:lemma2}
\end{lem}
Again this induces injective linear maps
\begin{eqnarray*}
  \iota_k: \Omega^2(\mathcal{M}) & \longrightarrow & \Omega^{2k+2}(\mathcal{M})
\\
  \alpha & \longmapsto & \alpha\wedge\omega^k
\end{eqnarray*}
for $k = 1,\ldots, n-2$. Thus we can obtain that
\begin{displaymath}
  \XS^1(\mathcal{M},\omega) = \XS^3(\mathcal{M},\omega) = \ldots
  = \XS^{2n-3}(\mathcal{M},\omega) \subseteq \XS^{2n-1}(\mathcal{M},\omega).
\end{displaymath}
When $k = n-2$, it can be easily verified from lemma
\ref{lem:lemma2} that
\begin{eqnarray}
  \iota = \iota_{n-2}: \Omega^2(\mathcal{M}) & \longrightarrow &
  \Omega^{2n-2}(\mathcal{M})
\nonumber \\
  \alpha & \longmapsto & \alpha\wedge\omega^{n-2}
\end{eqnarray}
is a linear isomorphism. If $n = 2$, we use the convention that
$\omega^0 = 1$. Namely, $\iota = \mathrm{id}: \alpha \longmapsto
\alpha$. Then we can define a linear map $\phi$ making the
following diagram commutative:
\begin{equation}
  \begin{CD}
    \Omega^2(\mathcal{M}) @>\iota>> \Omega^{2n-2}(\mathcal{M}) \\
    @V\phi VV                        @V\dd VV\\
    \XH^{2n-1}(\mathcal{M},\omega) @>\nu_n>> B^{2n-1}(\mathcal{M}).
  \end{CD}
\label{phicg}
\end{equation}

In general, it can be shown that there exist certain symplectic
manifolds on which the $2k-1$-st Euler-Lagrange cohomologe group
$H^{2k-1}_{\mathrm{EL}}(\mathcal{M},\omega), ~k\neq 1,n,$ is not
isomorphic to either de Rahm cohomology group
$H^{2k-1}_{\mathrm{dR}}(\mathcal{M})$ or the harmonic cohomology
group $H^{2k-1}_{\mathrm{har}}(\mathcal{M},\omega)$ on the
manifold \cite{gpwz2}. Therefore, the Euler-Lagrange cohomology
groups for $k\neq 1,n$ are new cohomological property of the
symplectic manifold in general.

\section{The General Form of Volume Preserving Equations}

Now we concentrate on the general form of volume preserving
equations from the cohomological point of view. We will first
present such kind of equations in the first subsection. Then we
illustrate some concrete volume-preserving systems. Especially, we
show that the ordinary canonical equations (\ref{Heqs}) are
included as a special case. Finally, we explore some of the
properties of important 2-forms for these volume-preserving
systems.

\subsection{The general volume preserving equations \omits{and
  $B^{2n-1}_{\rm{EL}}(\cal M)$}}

It is important to see that as shown in the diagram (\ref{phicg}),
for a given 2-form
\begin{equation}
  \alpha = \frac{1}{2}\,Q_{ij}\,\dd q^i\wedge\dd q^j
  + A^i_j\,\dd p_i\wedge\dd q^j + \frac{1}{2}\,P^{ij}\,\dd p_i\wedge\dd p_j
\label{alpha}
\end{equation}
where $Q_{ij}$ and $P^{ij}$ are antisymmetric for each pair of $i$ and $j$,
the vector field
\begin{displaymath}
  \phi(\alpha) = (\nu_n^{-1}\circ\dd\circ\iota)(\alpha)
  = \nu^{-1}_n(\dd\alpha\wedge\omega^{n-2})
\end{displaymath}
is in $\XH^{2n-1}(\mathcal{M},\omega)$.

For convenience, we set
\begin{displaymath}
  \vect{X} = n(n-1)\,\phi(\alpha)
  = n(n-1)\,\nu_n^{-1}(\dd\alpha\wedge\omega^{n-2}),
\end{displaymath}
i.e., $i_{\vect{X}}(\omega^n) = -
n(n-1)\,(\dd\alpha)\wedge\omega^{n-2}$. It is
easy to obtain that
\begin{equation}
  \vect{X} = \bigg(\frac{\partial P^{ij}}{\partial q^j}
  + \frac{\partial A^j_j}{\partial p_i} - \frac{\partial A^i_j}{\partial p_j}
  \bigg)\,\frac{\partial}{\partial q^i}
  + \bigg(\frac{\partial Q_{ij}}{\partial p_j}
  - \frac{\partial A^j_j}{\partial q^i}
  + \frac{\partial A^j_i}{\partial q^j}\bigg)\,
  \frac{\partial}{\partial p_i}.
\label{XH}
\end{equation}
 In fact,
\begin{eqnarray*}
  \dd\iota(\alpha)&=&\dd\alpha\wedge\omega^{n-2} \\
  &=& \frac{1}{2}\,\frac{\partial Q_{ij}}{\partial q^k}\,
  \dd q^i\wedge\dd q^j\wedge\dd q^k\wedge\omega^{n-2}
  +\frac{1}{2}\,\frac{P^{ij}}{\partial p_k}\,
    \dd p_i\wedge\dd p_j\wedge\dd p_k\wedge\omega^{n-2}
\\
  & & +\bigg(\frac{1}{2}\,\frac{\partial Q_{jk}}{\partial p_i}
    + \frac{\partial A^i_j}{\partial q^k}\bigg)\,
    \dd p_i\wedge\dd q^j\wedge\dd q^k\wedge\omega^{n-2}
\\
  & & + \bigg( \frac{1}{2}\,\frac{\partial P^{ij}}{\partial q^k}
    -\frac{\partial A^i_k}{\partial p_j} \bigg)\,
    \dd p_i\wedge\dd p_j\wedge\dd q^k\wedge\omega^{n-2}
\\
  &=&  \bigg(\frac{1}{2}\,\frac{\partial Q_{jk}}{\partial p_i}
    + \frac{\partial A^i_j}{\partial q^k}\bigg)\,
    \dd p_i\wedge\dd q^j\wedge\dd q^k\wedge\omega^{n-2}
\\
  & & + \bigg( \frac{1}{2}\,\frac{\partial P^{ij}}{\partial q^k}
    -\frac{\partial A^i_k}{\partial p_j} \bigg)\,
    \dd p_i\wedge\dd p_j\wedge\dd q^k\wedge\omega^{n-2}.
\end{eqnarray*}
By virtue of the following two equations
\begin{eqnarray}
  \dd p_i\wedge\dd p_j\wedge\dd q^k \wedge\omega^{n-2}
  & = & \frac{\delta^k_j}{n-1}\,\dd p_i \wedge\omega^{n-1}
  - \frac{\delta^k_i}{n-1}\,\dd p_j \wedge\omega^{n-1},
\\
  \dd p_i\wedge\dd q^j\wedge\dd q^k\wedge\omega^{n-2}
  & = & \frac{\delta^j_i}{n-1}\,\dd q^k\wedge\omega^{n-1}
  - \frac{\delta^k_i}{n-1}\,\dd q^j\wedge\omega^{n-1},
\end{eqnarray}
we can write $\dd\iota(\alpha)$ as
\begin{eqnarray*}
  \dd\iota(\alpha)
  & = & \frac{1}{n-1}\bigg(\frac{\partial A^j_j}{\partial q^i}
  - \frac{\partial A^j_i}{\partial q^j} - \frac{\partial Q_{ij}}{\partial p_j}
  \bigg)\,\dd q^i\wedge\omega^{n-1}
\nonumber \\ & &
  + \frac{1}{n-1}\bigg(\frac{\partial P^{ij}}{\partial q^j}
  + \frac{\partial A^j_j}{\partial p_i} - \frac{\partial A^i_j}{\partial p_j}
  \bigg)\,\dd p_i\wedge\omega^{n-1}
\\
  & = & \frac{1}{n(n-1)}\bigg(\frac{\partial A^j_j}{\partial q^i}
  - \frac{\partial A^j_i}{\partial q^j} - \frac{\partial Q_{ij}}{\partial p_j}
  \bigg)\,i_{\frac{\partial}{\partial p_i}}\omega^n
\nonumber \\ & &
  - \frac{1}{n(n-1)}\bigg(\frac{\partial P^{ij}}{\partial q^j}
  + \frac{\partial A^j_j}{\partial p_i} - \frac{\partial A^i_j}{\partial p_j}
  \bigg)\,i_{\frac{\partial}{\partial q^i}}\omega^n.
\end{eqnarray*}
Comparing it with
\begin{equation}
  \dd\iota(\alpha) = \frac{1}{n(n-1)}\,\nu_n(\vect{X})
  = -\frac{1}{n(n-1)}\,i_{\vect{X}}(\omega^n),
\label{nunX}
\end{equation}
we can obtain the expression of $\vect{X}$, as shown in eq.~(\ref{XH}).

Since both $\iota$ and $\nu_n$ are linear isomorphisms, we can see
from the commutative diagram (\ref{phicg}) that for each 2-form
$\alpha$ on $\mathcal{M}$ as in eq.~(\ref{alpha}) the vector field
$\vect{X}$ in eq.~(\ref{XH}) belongs to
$\XH^{2n-1}(\mathcal{M},\omega)$ and for each
  $\vect{X}\in\XH^{2n-1}(\mathcal{M},\omega)$
there exists the 2-form $\alpha$ on $\mathcal{M}$ satisfying
eqs.~(\ref{XH}). But, there may exist several 2-forms that are
mapped to the same vector field $\vect{X}$. For example, the
vector field $\vect{X}$ in (\ref{XH}) is invariant under the
transformation
\begin{equation}
  \alpha \longmapsto \alpha + \theta
\label{transf}
\end{equation}
where $\theta$ is a closed 2-form.

Note that for  $\vect{X}\in\XH^{2n-1}(\mathcal{M},\omega)$, the
2-form $\alpha$ is a globally defined differential form on
$\mathcal{M}$. If $\vect{X}\in\XS^{2n-1}(\mathcal{M},\omega)$,
such a 2-form cannot be found if
$H^{2n-1}_{\mathrm{dR}}(\mathcal{M})$ is nontrivial. In this case,
$\alpha$ can  still be found as a locally defined 2-form,
according to the Poincar\'e lemma. Then the relation between
$\vect{X}$ and the locally defined 2-form $\alpha$,
eq.~(\ref{XH}), is valid only on a certain  open subset of
$\mathcal{M}$; And on the intersection of two such open subsets,
the corresponding 2-forms are not identical. That is, the
transformation relation of $Q_{ij}$, $A^i_j$ as well as $P^{ij}$
is not that of tensors on $\mathcal{M}$. Instead, a transformation
like (\ref{transf}) or more complicated should be applied.

No matter whether $\vect{X}$ belongs to
$\XH^{2n-1}(\mathcal{M},\omega)$ or
$\XS^{2n-1}(\mathcal{M},\omega)$, i.e., whether the 2-form
$\alpha$ is globally or locally defined, the flow of $\vect{X}$
can be always obtained provided that the general solution of the
following equations can be solved:
\begin{eqnarray}
  \dot{q}^{\,i} & = & \frac{\partial P^{ij}}{\partial q^j}
  + \frac{\partial A^j_j}{\partial p_i}
  - \frac{\partial A^i_j}{\partial p_j},
\nonumber \\
  \dot{p}_i & = & \frac{\partial Q_{ij}}{\partial p_j}
  - \frac{\partial A^j_j}{\partial q^i}
  + \frac{\partial A^j_i}{\partial q^j}.
\label{evps}
\end{eqnarray}
This is just the general form of the equations of
a volume-preserving mechanical system on a
symplectic manifold $(\mathcal{M},\omega)$.


\subsection{Some concrete volume-preserving
systems.}

Let us show some concrete volume-preserving systems.

First, since the canonical equations (\ref{Heqs}) in the
Hamiltonian mechanics is volume preserving, it should be a special
case of (\ref{evps}). In fact, this is just the case.

For the Hamiltonian $H$ of the canonical system on $\mathcal{M}$,
take the 2-form as
\begin{equation}
  \alpha = \frac{1}{n-1}\,H\,\omega,
\end{equation}
then eqs.~(\ref{evps}) just turn out to be the canonical equations
(\ref{Heqs}) as it should be. \omits{Thus the ordinary Hamilton
equations have been included as a special case, as what is
expected.}

Secondly, let us recall the kind of linear systems mentioned in
\S\ref{sect:Intro} on $(\mathbb{R}^{2n},\omega)$ with $\omega=\dd p_i
\wedge\dd q^i$.

Taking
  $Q_{ij} = - a_{ij}\,p_k q^k$, $A^i_j = \frac{1}{n-1}\,H\,\delta^i_j$
and $P^{ij} = 0$
with the constants $a_{ij}$ and the function $H$ as shown in eqs.~(\ref{lsysH})
and (\ref{lsyscoef}), namely,
\begin{equation}
  \alpha = \frac{1}{n-1}\,H\,\omega
  - \frac{1}{2}\,p_k q^k\,a_{ij}\,
  \dd q^i\wedge\dd q^j,
\label{lsysalpha}
\end{equation}
then the eqs (\ref{evps})  turn out to be eqs.~(\ref{lsys}).
Therefore the 2-form $\alpha$ in eq.~(\ref{lsysalpha}) is one of
general forms corresponding to such kind of linear systems. But,
it is not the unique.

It is interesting that when $H = 0$ in the above equation, all the coordinates
$q^i$ are the first integrals of the linear system. Then all the canonical
momenta
  $p_i = a_{ij}\,q^j t + p_{i0}$ where $p_{i0}$
are constants. This can generalize to an arbitrary symplectic
manifold for the equations (\ref{evps}), even though the system is
no longer a linear system.

Thirdly, let us now consider some kind of volume preserving
systems on a generic symplectic manifold $(\mathcal{M},\omega)$.

If the 2-form $\alpha$ satisfies on a Darboux coordinate
neighborhood $(U; q,p)$ in $\mathcal{M}$
\begin{equation}
  i_{\frac{\partial}{\partial p_i}}\alpha
  = A^i_j\,\dd q^j + P^{ij}\,\dd p_j
  = 0
\end{equation}
for each $i=1,\ldots,n$, then $A^i_j = 0$ and $P^{ij} = 0$. Hence, on that
coordinate neighborhood $U$, all the $q^i$ are constant.


\subsection{The trace of 2-forms}

If we define a function $\tr\alpha$ as
\begin{equation}
  \alpha\wedge\omega^{n-1} = \frac{\tr\alpha}{n} \,\omega^n
\end{equation}
for each 2-form $\alpha$, then using the formula
\begin{displaymath}
  \dd p_i\wedge\dd q^j\wedge\omega^{n-1} = \frac{\delta^j_i}{n}\,\omega^n,
\end{displaymath}
we obtain that
\begin{equation}
  \tr\alpha = A^i_i.
\end{equation}
The above expression is obviously independent of the choice of
the Darboux
coordinates.

Let $\vect{X}_{\tr\alpha}$ be the Hamiltonian vector field
corresponding to the function $\tr\alpha$. Eq.~(\ref{XH})
indicates that
\begin{equation}
  \vect{X} = \vect{X}_{\tr\alpha} + \vect{X}'
\end{equation}
where $\vect{X}'$ is the extra part on the right hand side of
eq.~(\ref{XH}), corresponding to the 2-form
\begin{equation}
  \alpha - \frac{\tr\alpha}{n - 1}\,\omega.
\end{equation}
\comments{  
Hence, when the 2-form $\alpha = \frac{H}{n-1}\,\omega$ with $H$
a function on
$\mathcal{M}$, $\tr\alpha = \frac{n}{n-1}\,H$ and the traceless
part of
$\alpha$ vanishes.
}  

If $f(q,p)$ is a function on $\mathcal{M}$, then the derivative
  $\dot{f} = \frac{\dd}{\dd t}f(q(t),p(t))$
along any one of the integral curves of
eqs.~(\ref{evps}) satisfies the equation
\begin{equation}
  \dot{f}\,\omega^n
    = n(n-1)\,\dd\alpha\wedge\dd f\wedge\omega^{n-2}.
\label{dotf}
\end{equation}
In fact, $\dot{f}(t) = (\Lied{X}f)(q(t),p(t))$.
And, since $\vect{X}$ is volume-preserving,
\begin{displaymath}
  (\Lied{X}f)\,\omega^n = \Lied{X}(f\,\omega^n)
  = \dd i_{\vect{X}}(f\,\omega^n) + i_{\vect{X}}\dd(f\,\omega^n)
  = \dd (f\,i_{\vect{X}}\omega^n).
\end{displaymath}
Then according to eq.~(\ref{nunX}),
\begin{eqnarray*}
  (\Lied{X}f)\,\omega^n & = & - n(n-1)\,\dd(f\,\dd\iota(\alpha))
  = - n(n-1)\,\dd f\wedge\dd(\alpha\wedge\omega^{n-2})
\\
  & = & - n(n-1)\,\dd f\wedge\dd\alpha\wedge\omega^{n-2}
  = n(n-1)\,\dd\alpha\wedge\dd f\wedge\omega^{n-2}.
\end{eqnarray*}
Thus eq.~(\ref{dotf}) has been proved.

Especially, when $\alpha = \frac{H}{n-1}\,\omega$, the system
(\ref{evps})
turns out to be the usual Hamiltonian system, as we have
mentioned. For such
a Hamiltonian system, on the one hand, we can use eq.~(\ref{dotf})
to obtain
\begin{displaymath}
  \dot{f}\,\omega^n
  = n\,\dd(H\,\omega)\wedge\dd f\wedge\omega^{n-2}
  = n\,\dd H\wedge\dd f\wedge\omega^{n-1}
  = \tr(\dd H\wedge\dd f)\,\omega^n,
\end{displaymath}
namely,
\begin{equation}
  \dot{f} = \tr(\dd H\wedge\dd f).
\end{equation}
On the other hand, $\dot{f}$ can be expressed in terms of the
Poisson bracket:
\begin{displaymath}
  \dot{f} = \{f,H\} := \vect{X}_H f
  = \frac{\partial f}{\partial q^i}\frac{\partial H}{\partial
  p_i}
  - \frac{\partial f}{\partial p_i}\frac{\partial H}{\partial q^i}.
\end{displaymath}
So we obtain the relation between the Poisson bracket and the
trace of 2-forms:
\begin{equation}
  \{f,H\} = - \tr(\dd f\wedge\dd H).
\end{equation}


\section{The Relations with Other
Volume Preserving Systems}

Let us now consider the relations between our volume-preserving
equations and other relevant topics such as Feng-Shang's volume
preserving systems and the Nambu mechanics.

\subsection{Feng-Shang's volume preserving systems}

In order to develop the volume-preserving algorithm, Feng and
Shang \cite{FS} presented the following lemma.

\begin{thm}[Feng-Shang's lemma]
The volume-preserving vector field
   $X = (X^1, \ldots, X^n)^T$
on $\mathbb{R}^n$ can always be expressed by an antisymmetric
tensor $a^{ij}$ on $\mathbb{R}^n$ as
\begin{equation}
  X^i = \frac{\partial a^{ij}}{\partial x^j},
\label{divfree}
\end{equation}
where $x^i$ are the standard coordinates on $\mathbb{R}^n$.
\end{thm}

Note that here we have adopted the notations so as to accommodate
ours in this paper. In fact, this lemma can be  understood as
follows: A vector field on $\mathbb{R}^n$, known as a Euclidean
space or a Riemannian manifold, is volume-preserving if and only
if it is divergence-free. Using the Hodge theory, the divergence
of the vector field $X$ reads
  $\mathrm{div} X = -\delta\tilde{X} = *\ \dd *\tilde{X}$,
where $\delta$ is the codifferential operator and
  $\tilde{X} = \delta_{ij}\,X^i\,\dd x^j$
is the 1-form (covariant vector field) obtained from $X$ by
lowering indices with the metric $\delta_{ij}$ for $\mathbb{R}^n$.
When $X$ is volume-preserving, $\mathrm{div} X = *\ \dd *\tilde{X} =
0$. This implies that $*\tilde{X}$ is a closed $(n-1)$-form. Then,
according to the Poincar\'e lemma, there is a 2-form $\alpha$,
say, such that $*\tilde{X} = \dd *\alpha$. If $\alpha$ is assumed
to be $\frac{1}{2}\,a_{ij}\,\dd x^i\wedge\dd x^j$ with $a_{ij} = -
a_{ji} = \delta_{ik}\delta_{jl}\,a^{kl} = a^{ij}$, we can obtain
that $\tilde{X} = (-1)^{n-1}\,*\dd *\alpha
   = \delta_{ij}\,\frac{\partial a^{jk}}{\partial x^k}\,\dd x^i$.
Then the vector field $X$ satisfies eq.~(\ref{divfree}).

According to the precise sense of volume, Feng-Shang's lemma can
be grouped into the same class with the approach to the
volume-preserving systems proposed by Nambu \cite{Nambu}. This
lemma has presented the most general form of volume-preserving
systems in this class.

It is worth seeing that the formula ~(\ref{divfree}) is quite
similar to eq.~(\ref{XH}), only with some slight differences: (1)
This formula is a statement on the Euclidean space $\mathbb{R}^n$
with an arbitrary dimension $n$ while eq.~(\ref{XH}) is for a
symplectic manifold, of $2n$-dimensional. (2) It can be
generalized to an $n$-dimensional Riemannian or pseudo-Riemannian
manifold provided that the $(n-1)$-st de Rham cohomology group is
trivial, with the ordinary derivatives being replaced by the
covariant derivatives. As for eq.~(\ref{XH}), when
$H^{2n-1}_{\mathrm{dR}}(\mathcal{M}) \neq 0$, not every
volume-preserving vector field satisfies it. This is similar to
eq.~(\ref{divfree}). But a covariant derivative is not necessary
in eq.~(\ref{XH}). (3) When the symplectic manifold $\mathcal{M} =
\mathbb{R}^{2n}$, we can set
\begin{displaymath}
  (a_{ij})_{2n\times 2n}
  = \left( \begin{array}{rr} 0 &  I \\ - I & 0 \end{array} \right)
    \left( \begin{array}{rc} Q & -A^T \\ A & P \end{array} \right)
    \left( \begin{array}{rr} 0 & - I \\ I & 0 \end{array} \right)
  = \left(
  \begin{array}{lr}
    P   & -A \\
    A^T & Q
  \end{array}
  \right)
\end{displaymath}
where $Q = (Q_{ij})_{n\times n}$, $P = (P^{ij})_{n\times n}$,
$A=(A^i_j)_{n\times n}$ and $I$ is the $n\times n$ unit matrix.
Accordingly, we set
  $(x^1,\ldots,x^n,x^{n+1},\ldots,x^{2n}) = (q^1,\ldots,q^n,p_1,\ldots,p_n)$.
Then eq.~(\ref{divfree}) reads
\begin{equation}
  X = \bigg(
   \frac{\partial P^{ij}}{\partial q^j} - \frac{\partial A^i_j}{\partial p_j},
   \frac{\partial A^j_i}{\partial q^j} + \frac{\partial Q_{ij}}{\partial p_j}
  \bigg)^T,
\end{equation}
with the index $i$ running form 1 to $n$. Comparing it with
eq.~(\ref{XH}), it
seems that a condition
\begin{equation}
  \tr \alpha = 0
\end{equation}
could be imposed on the 2-form $\alpha$ in eq.~(\ref{XH}).

\subsection{The Nambu mechanics}

It should be mentioned that the idea of generalizing the
Hamiltonian systems can be traced back to Nambu \cite{Nambu}. In
\cite{Nambu}, Nambu pointed out that there are various
possibilities of generalizing the Hamiltonian systems to the
volume-preserving systems. Especially he discussed in detail the
volume-preserving system on $\mathbb{R}^{3N}$. His discussion can
smoothly generalizes to a $3N$-manifold $M_1\times
M_2\times\ldots\times M_N$, where $M_a$ for each $a$ is a
3-dimensional Riemannian (or pseudo-Riemannian) manifold. The most
exciting is that the quantization can be achieved within such a
mechanics. --- The ordinary quantization of mechanics, including
the geometric quantization \cite{GQ}, is based on the
Hamiltonian mechanics. 

Our approach presented in this paper, however, differs from
Nambu's work in three aspects. First, our interest  is the
volume-preserving systems on a symplectic manifold. We are not
about to generalize the Hamiltonian systems, although certain a
generalization has been inevitably resulted in. Secondly, the
systems in Nambu's approach preserve a volume induced by the
metric on that manifold. In Nambu's paper, in order to perform the
vector product, the metric is necessary. The volume preserved in
our approach is induced by the symplectic structure. So, such a
generalization is  different from that of Nambu. Thirdly, as
having been discussed, the ordinary Hamiltonian systems have been
included as the special cases into our general form of
volume-preserving equations. This is not the case in Nambu's
approach.

\section{Conclusion and Remarks}

In this paper, we have briefly introduced the definition and
properties of the Euler-Lagrange cohomology groups on a symplectic
manifold $(\mathcal{M, \omega})$ and presented the general form of
equations that relate to the image of the highest Euler-Lagrange
cohomology group, which generate a volume-preserving flow on the
manifold. It has been shown that for every volume-preserving flow
there are
 some 2-forms playing the role similar to the Hamiltonian functions in
the Hamilton mechanics and the ordinary canonical equations with
Hamiltonian $H$ are included as a special case with the 2-form
$\frac{1}{n-1}\,H\,\omega$. Thus, this is a generalization of the
Hamilton mechanics from the cohomological point of view.\omits{
where $H$ is the corresponding Hamiltonian.}

 Finally, it should
be mentioned that the volume-preserving systems play very
important role in  physics, especially in the both classical
mechanics and statistical physics via the famous Liouville
theorems in both of them. However, this has not been noticed widely
yet. In this paper we have not explained why the volume-preserving
systems are so important in statistical physics, but left it as a
topic for the forthcoming papers.

\section*{Acknowledgement}

We would like to thank Professor Z. J. Shang, S. K. Wang and Siye
Wu for valuable discussions. This work was supported in part by
the National Natural Science Foundation of China (grant Nos.
90103004, 10171096, 19701032, 10071087) and the National Key
Project  for Basic Research of China (G1998030601).

\end{document}